\documentclass[12pt]{article}
\usepackage{graphicx}
\usepackage{cite}
\usepackage{amsmath,amssymb}
\usepackage{bm}
\usepackage{subcaption}
\usepackage{caption}
\usepackage{xcolor,soul,tikz}
\usepackage{hyperref}
\usepackage[top=2.2cm,bottom=2.2cm,left=1.8cm,right=1.8cm]{geometry}

\title{Eddy currents and error fields in the toroidally segmented shell configuration aka Armadillo}

\author{
	D. Abate\textsuperscript{1}\thanks{Corresponding author: domenico.abate@igi.cnr.it} 
	\and
		A. Corbioli\textsuperscript{1, 2} \and
	G. Marchiori\textsuperscript{1} 
}
	
\begin{document}
\maketitle

	\begin{center}
	\textsuperscript{1}	Consorzio RFX (CNR, ENEA, INFN, Università di Padova,\\ Acciaierie Venete SpA) \\
	\textsuperscript{2}CRF – Università degli Studi di Padova, Padova, Italy\\
\end{center}

\begin{center}
	\textit{To be submitted to Plasma Physics and Controlled Fusion.}
\end{center}

\begin{abstract}
	A stream function formulation for the calculation of the eddy current distribution in a
	toroidally segmented conducting shell (Armadillo) is presented and
	verified. The governing Poisson-like equation is solved spectrally
	via a Discrete Sine Transform, which enforces the zero-current
	boundary condition at each gap exactly by construction, and the
	related radial magnetic field at the plasma surface is obtained by
	Biot-Savart integration. The toroidal eddy current flows in the same
	direction on every sector with a cosine-like poloidal distribution, while
	the poloidal return current, required by current continuity, peaks
	near each gap and reverses sign between adjacent sectors. The scaling
	of both components with the number of gaps is governed by the
	competition between the poloidal geometric scale of the shell and the
	toroidal segmentation scale. The transition between the two regimes is set
	by the aspect ratio of the device, around eight gaps for an
	RFX-mod2-like geometry, as predicted by the analytical wall time
	model of the Armadillo configuration. At the gap, the poloidal return current
	produces a radial field of opposite sign to the value far from the gap; the
	resulting non-axisymmetric error field peaks near this transition and
	becomes negligible for highly segmented shells.  These results provide a first characterisation of
	the eddy currents and error fields of the Armadillo configuration,
	needed for future resistive wall mode and tearing mode stability
	analyses.
\end{abstract}

\section{Introduction}
In magnetic confinement fusion devices, the conducting structures
surrounding the plasma carry eddy currents during transient phases of
the discharge. When induced by a growing magnetohydrodynamic (MHD)
perturbation, these currents, whether axisymmetric or not, oppose the
change of magnetic flux through the structure and provide passive
stabilization. Eddy currents are also induced by the evolution of the
equilibrium, e.g.\ during the plasma current ramp-up, and their
distribution becomes non-axisymmetric when the conducting path is
interrupted by gaps, welds or joints~\cite{lazarus1990control}.
Together with coil misalignments and ferromagnetic components, these
structural asymmetries produce non-axisymmetric magnetic
perturbations, collectively referred to as error fields, which
perturb the plasma equilibrium and degrade confinement in both tokamak
and reversed field pinch plasmas~\cite{strait2003resistive,
	boozer2004physics, fitzpatrick1995helical}. The effect of an error
field is not given by its vacuum amplitude alone since the plasma response
can either amplify or reduce it depending on its coupling to
the actual plasma response~\cite{park2008error, logan2020empirical}.
In reversed field pinches (RFPs), a toroidal passive stabilizing
shell (PSS) is used to slow MHD instabilities from the ideal Alfvén timescale to a resistive one,
enabling active feedback control~\cite{pustovitov2008general,
	marchiori2007characterization}. The presence of cuts in the PSS is
necessary to allow penetration of the magnetic fields produced by the
external coils. As a consequence, the same eddy currents responsible
for passive stabilization also generate error fields due to the
non-uniform current distribution near structural discontinuities such
as poloidal and toroidal gaps. In RFX-mod, the error fields associated
with the PSS gaps have been quantified using both analytical \cite{marchioritn,masiello_analytical_1998} and numerical 3D electromagnetic
models~\cite{bettini2017modeling, marrelli2019optimization}, showing
the need for active compensation during current ramp-up. 
An improvement in RFP confinement could be achieved by increasing the
plasma–shell proximity, removing the highly resistive vacuum vessel and
placing the PSS directly in vacuum inside a vacuum-tight toroidal
support structure. This is the main change from RFX-mod to the upgrade
RFX-mod2 	\cite{marrelli2019upgrades}. In such a configuration, stringent electrical insulation
requirements arise because of diffuse plasma surrounding the PSS,
requiring a full alumina coating~\cite{cordaro2024electrical,
	cordaro2023electrical}.

A more efficient alternative is the toroidally segmented shell, known
as the Armadillo configuration \cite{walltimearmadillo2026}, conceived during the RFX-mod2 design
process. Introducing
$N_g$ toroidal gaps reduces the induced electromotive force on each
segment by a factor $1/N_g$, relaxing insulation requirements. The
impact of segmentation on eddy-current dynamics was first studied
analytically in terms of wall time — the characteristic timescale for
magnetic field diffusion through a conducting structure, which governs
both passive stabilization and the bandwidth required for feedback
control~\cite{strait2003resistive, chu2010stabilization} — for the
axisymmetric $(m=1,\,n=0)$ mode~\cite{nagayama1984feedback}. The
dependence of the wall time on geometry and mode numbers has since been
studied for continuous shells~\cite{gimblett1986free,
	pustovitov2008general, pustovitov2018reaction, chukashev2025toroidal},
and the distinction between different characteristic wall times has
been clarified in~\cite{dialetis1991diffusion, pustovitov2008general,
	pustovitov2018reaction}. More recently, a general analytical expression
for the wall time of a general helicity magnetic field diffusing through an exact toroidally segmented shell (Armadillo) has been derived~\cite{walltimearmadillo2026},
showing that segmentation introduces a quadratic correction to the
effective wall resistivity, yielding $\tau_w\propto N_g^{-2}$ in the
strongly segmented regime. While this model describes the temporal
decay of eddy currents, it does not provide their spatial distribution
or the resulting error fields at the plasma boundary. The investigation
of the magnitude and spatial structure of the eddy currents in the
Armadillo configuration is crucial since it has direct implications for
MHD stability, affecting both the resistive wall modes — governed by
$\tau_w$ — and, particularly, the tearing mode dynamics, which are
affected by the error fields~\cite{fitzpatrick1993interaction, zanca2009avoidance,
	zanca2025phase}.

In this work, we present a stream function formulation for the calculation of the eddy-current distribution in a toroidally segmented shell.
The Poisson-like equation for the stream function is solved spectrally using a Discrete Sine Transform (DST) in the toroidal direction, which
enforces the boundary condition on current exactly by
construction. The method is applied to the RFX-mod2 geometry for a
typical ramp-up phase of a 2\,MA plasma using RFX-mod experimental data (shot \#29262 at
$t=20\,\mathrm{ms}$), and the results are in accordance with previous calculations \cite{eps2026} performed using the 3D electromagnetic code CARIDDI~\cite{Albanese_CARIDDI}. The error fields generated for different numbers of gaps are quantified and analyzed.

\section{Circuital model}
\label{sec:circuit}
\begin{figure}[tb!]
	\centering
	\includegraphics[width=\textwidth]{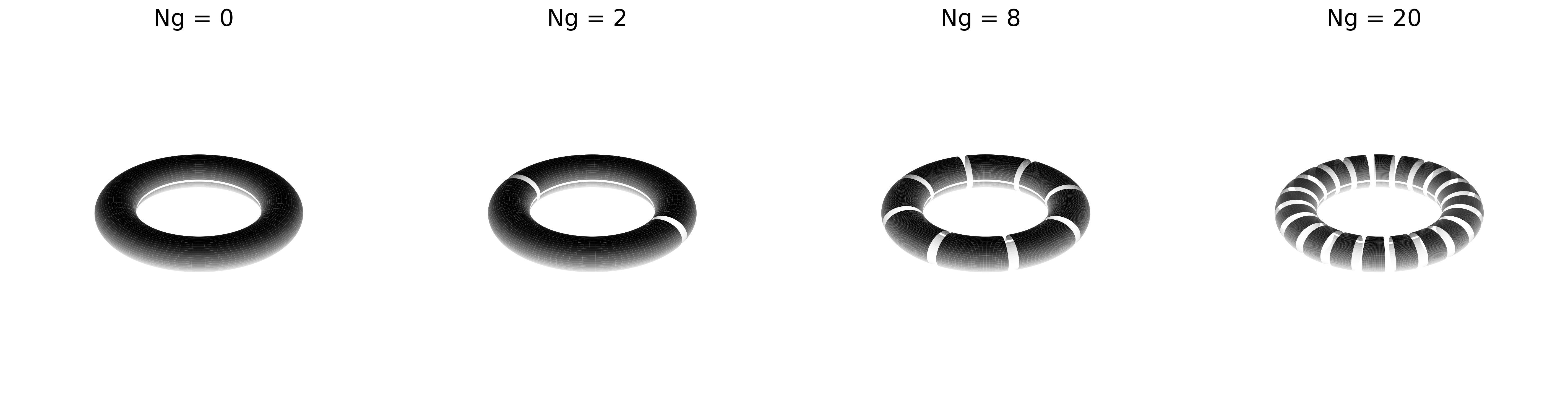}
	\caption{Three-dimensional representation of the segmented toroidal
		shell for $N_g=0$ (continuous), $2$, $8$ and $20$ toroidal
		gaps. The poloidal gap at the inboard midplane ($\theta=\pi$)
		is present in all configurations.}
	\label{fig:shell_geom}
\end{figure}

The Armadillo configuration consists of a toroidal conducting shell
divided into $N_g$ identical sectors by equally spaced toroidal gaps
— interruptions of the conducting path at fixed toroidal angles
$\phi_j = 2\pi j/N_g$, $j=0,\ldots,N_g-1$, enforcing $J_\phi=0$ at
each gap location. The resulting segmented shell
geometry is illustrated in Fig.~\ref{fig:shell_geom}.
 In general, a poloidal gap may also be present: this is an interruption at a fixed poloidal angle $\theta=\theta_g$ enforcing
$J_\theta=0$. In this work we consider the geometry of RFX-mod2: a toroidal copper
shell of major radius $R_0=1.995\,\mathrm{m}$, minor radius
$r_w=0.5115\,\mathrm{m}$, thickness $t_w=3\,\mathrm{mm}$, and
electrical resistivity $\eta=0.018\,\mu\Omega\mathrm{m}$, with a
poloidal gap fixed at the inboard midplane ($\theta_g=\pi$) and two toroidal gaps $N_g=2$. The number of toroidal gaps $N_g$ is also varied systematically in order to characterize the
Armadillo configuration.

The modelling approach proceeds in three steps. First, the mutual
inductances between the plasma current, the field shaping coils, and
the shell are computed via the Neumann formula~\cite{neumann}, and
the total induced voltage $V_\phi^\mathrm{tot}$ on the shell is
evaluated from the measured current time derivatives. Second, the $N_g$ toroidal gaps are
introduced analytically: the total induced voltage is distributed
over $N_g$ identical sectors so that the voltage per sector is
$V_\phi^\mathrm{sector}=V_\phi^\mathrm{tot}/N_g$ and the driving
electric field:
\begin{equation}
	E_\phi = \frac{V_\phi^\mathrm{tot}}{2\pi R_0}
\end{equation}
is independent of $N_g$. Third, the eddy current distribution is
computed via a stream function formulation with a spectral DST solver
(Sec.~\ref{sec:eddy}), and the magnetic field at the plasma surface
is evaluated via the Biot-Savart law.

The induced electromotive force on the shell follows Faraday's law:
\begin{equation}
	V = -\frac{d\Phi}{dt}\,,
\end{equation}
where $\Phi$ is the magnetic flux. Modelling the shell as two separate
circuits — one toroidal and one poloidal — the flux in each circuit is
related to the currents in the other circuits via the mutual
inductance, $\Phi = M\,I$, so the induced voltage becomes
\begin{equation}
	V = -M\,\frac{dI}{dt}\,.
\end{equation}

Three active circuits contribute to the induced voltage on the shell:
the ohmic heating (OH) winding, which drives the poloidal flux swing
for plasma current ramp-up; the field shaping (FS) windings, which
control the horizontal equilibrium position of the plasma; and the
plasma itself. The induced voltage per sector is:
\begin{align}
	V_\theta^\mathrm{sector} &= -\frac{1}{N_g}\Bigl(
	M_{\text{OH-shell},\,\theta}\,\frac{dI_{\text{OH}}}{dt}
	+ M_{\text{FS-shell},\,\theta}\,\frac{dI_{\text{FS}}}{dt}
	+ M_{\text{pla-shell},\,\theta}\,\frac{dI_{\text{pla}}}{dt}
	\Bigr), \\
	V_\phi^\mathrm{sector}   &= -\frac{1}{N_g}\Bigl(
	M_{\text{OH-shell},\,\phi}\,\frac{dI_{\text{OH}}}{dt}
	+ M_{\text{FS-shell},\,\phi}\,\frac{dI_{\text{FS}}}{dt}
	+ M_{\text{pla-shell},\,\phi}\,\frac{dI_{\text{pla}}}{dt}
	\Bigr).
	\label{eq:voltages}
\end{align}
The inclusion of the OH winding is essential: during ramp-up it drives
a poloidal flux swing of $15\,\mathrm{Wb}$ \cite{stella1995rfx}, and its omission would
lead to a significant underestimate of the induced voltage.

The mutual inductance between two circuits is computed via the Neumann
formula~\cite{neumann}:
\begin{equation}
	M = \frac{\mu_0}{4\pi} \oint_{C_1} \oint_{C_2}
	\frac{d\mathbf{l}_1 \cdot d\mathbf{l}_2}{|\mathbf{r}_1 - \mathbf{r}_2|}\,,
\end{equation}
where the integrals are taken over the two circuit paths $C_1$ and
$C_2$, discretized as sets of filamentary toroidal loops. 
Only the toroidal direction is relevant
here: the poloidal coupling coefficients vanish by axisymmetry, or
are negligible compared to the toroidal ones.


The mutual inductance coefficients are computed individually for each
circuit element, since the current time derivatives and the geometric
coupling vary significantly across the winding. For the FS winding,
the mutual inductance of each of the 8 sector pairs (F1--F8, each
consisting of two coils A+B in series) increases monotonically from
the innermost to the outermost coil. The factor $\sim 2.4$ variation
between F1 and F8 confirms that a single average coefficient would
introduce a significant error. The induced voltage from the FS winding
is:
\begin{equation}
	V_\phi^{\mathrm{FS}} = -\sum_{k=1}^{8} M_{Fk,\phi}\,\frac{dI_{Fk}}{dt}
\end{equation}
where $M_{Fk,\phi}$ is the mutual inductance between the $k$-th FS
sector pair (two coils A+B in series) and the shell in the toroidal
direction, and $dI_{Fk}/dt$ is the measured current derivative of
that pair. The factor of 2 from the series connection is absorbed
into $M_{Fk,\phi}$.

For the OH winding, the four sectors show good symmetry
($M_\phi$ varies from $1.33$ to $1.88\,\mu\mathrm{H}$), justifying
the use of an average coefficient:
\begin{align*}
	\bar{M}_{\text{OH-shell},\,\phi}  &= 1.568\times10^{-6}\,\mathrm{H}
	\quad (N_\mathrm{OH}=40\text{ coils}),\\
	\bar{M}_{\text{pla-shell},\,\phi} &= 3.863\times10^{-6}\,\mathrm{H}.
\end{align*}

\begin{figure}[tb!]
	\centering
	\includegraphics[width=1\textwidth]{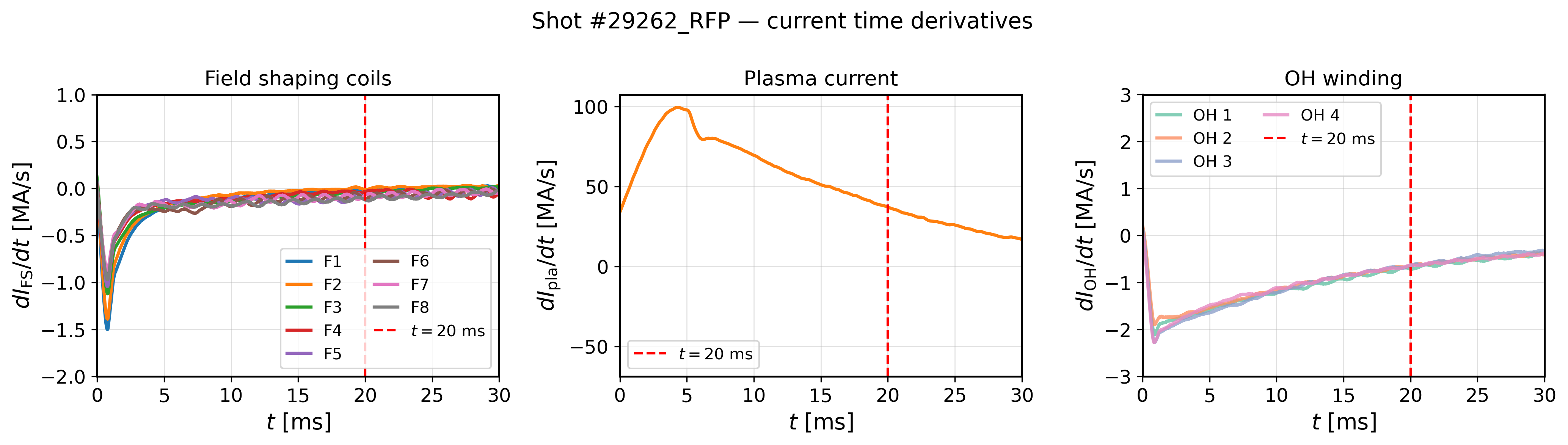}
	\caption{Time derivatives of the FS coil currents (left), plasma
			current (center) and OH winding (right) during shot \#29262.
			The vertical red line marks $t=20\,\mathrm{ms}$.}
	\label{fig:rampup_derivatives}
\end{figure}

During the ramp-up phase of shot \#29262 at $t=20\,\mathrm{ms}$,
the current derivatives are $d\bar{I}_\mathrm{OH}/dt =
-6.54\times10^5\,\mathrm{A/s}$ (average of 4 sectors, variation
$<14\%$) and $dI_\mathrm{pla}/dt = 3.72\times10^7\,\mathrm{A/s}$ as shown in Fig.~\ref{fig:rampup_derivatives}.
The FS derivatives vary significantly across sectors (std $=63\%$ of
mean), ranging from $-9.5\times10^3\,\mathrm{A/s}$ (F2) to
$-1.21\times10^5\,\mathrm{A/s}$ (F8). The corresponding voltage
contributions are summarized in Tab.~\ref{tab:contributions}: the
plasma term dominates ($-143.5\,\mathrm{V}$), followed by OH
($+41.0\,\mathrm{V}$) and FS ($+3.5\,\mathrm{V}$, with F8
contributing the most at $+0.96\,\mathrm{V}$ due to its larger mutual
inductance). The net total is $V_\phi^\mathrm{tot} \approx
-99.0\,\mathrm{V}$.

\begin{table}[tb]
	\centering
	\caption{Contributions to $V_\phi$ at $t=20\,\mathrm{ms}$
			(RFX-mod shot \#29262).}
	\label{tab:contributions}
	\begin{tabular}{l|c|c|c}
			Circuit & $N_\mathrm{coils}$ & effective $N\,d\bar{I}/dt$ (A/s)
			& $V_\phi^\mathrm{tot}$ (V) \\
			OH winding  & 40 & $-2.62\times10^{7}$ & $+41.0$ \\
			FS windings & 16 & $-1.06\times10^{6}$ & $+3.5$  \\
			Plasma      &  1 & $+3.72\times10^{7}$ & $-143.5$ \\ \hline
			\textbf{Total} & & & $\mathbf{-99.0}$ \\
		\end{tabular}
\end{table}

The toroidal electric field driving the induced currents is
$E_\phi = V_\phi^\mathrm{tot}/(2\pi R_0)$, which is independent of
$N_\mathrm{g}$: the factor $1/N_\mathrm{g}$ in the voltage per
sector cancels exactly with the factor $1/N_\mathrm{g}$ in the
sector arc length.

\section{Eddy currents calculation}
\label{sec:eddy}

The shell surface is described by the coordinates $(\theta,\phi)$,
with $\phi$ increasing counterclockwise when viewed from above and
$\theta$ measured from the outboard midplane, so that
$(\hat{\mathbf e}_r,\hat{\mathbf e}_\phi,\hat{\mathbf e}_\theta)$
is a right-handed triad with $\hat{\mathbf e}_r$ the outward normal.
In the thin-wall approximation ($t_w\ll r_w$), the shell carries a
surface current density $\mathbf{J}$ [A/m] with only poloidal
($J_\theta$) and toroidal ($J_\phi$) components. The condition
$\nabla\cdot\mathbf{J}=0$ is automatically satisfied by a scalar
stream function $\chi$ [A], $\mathbf{J}=\hat{\mathbf e}_r\times\nabla\chi$,
giving
\begin{equation}
	J_\theta = \frac{1}{h_\phi}\frac{\partial\chi}{\partial\phi}\,,
	\qquad
	J_\phi = -\frac{1}{h_\theta}\frac{\partial\chi}{\partial\theta}\,,
	\label{eq:J_from_chi_def}
\end{equation}
with $h_\theta=r_w$ and $h_\phi=R_0+r_w\cos\theta$. Taking the normal
component of the curl of Ohm's law, $\mathbf{J}=\sigma t_w\mathbf{E}$,
the stream function satisfies the Poisson equation
\begin{equation}
	\nabla^2\chi = f\,,
	\qquad
f = -\sigma t_w\,\hat{\mathbf e}_r\cdot\nabla\times\mathbf E
= \frac{\sigma t_w}{h_\theta h_\phi}
\bigl[\partial_\theta(h_\phi E_\phi)-\partial_\phi(h_\theta E_\theta)\bigr]
	\label{eq:poisson}
\end{equation}

Since the shell consists of $N_g$ identical sectors,
\eqref{eq:poisson} is solved on a single sector $\phi'\in[0,L_s]$,
where $L_s=2\pi/N_g$ is the angular width of the sector and $R_0L_s$
its physical arc length, and the full-torus solution is obtained by
repetition. In the cylindrical approximation $h_\phi\approx R_0$, the Laplacian
reads $\nabla^2\chi\approx r_w^{-2}\partial_\theta^2\chi
+R_0^{-2}\partial_{\phi'}^2\chi$. Its eigenfunctions
\begin{equation}
	u_{m,n}(\theta,\phi') = \sin(m\theta)\,
	\sin\!\left(\frac{n\pi\phi'}{L_s}\right)
	\label{eq:eigenfunctions}
\end{equation}
vanish at both toroidal gaps ($\phi'=0,L_s$) and at the poloidal gap
($\theta=\pi$), and satisfy
$\nabla^2 u_{m,n} = -\left(k_m^2+k_n^2\right)u_{m,n}$, with
\begin{equation}
	k_m = \frac{m}{r_w}\,,\qquad
	k_n = \frac{n\pi}{L_s R_0} = n\,k_s\,,\qquad
	k_s = \frac{N_g}{2R_0}\,,
	\label{eq:wavenumbers}
\end{equation}
where $k_s$ is the toroidal segmentation
wavenumber~\cite{walltimearmadillo2026}. The toroidal basis is the Discrete Sine Transform
(DST-I)~\cite{britanak2010discrete}. Expanding both the stream
function and the source on this basis,
\begin{equation}
	\chi = \sum_{m,n}\chi_{m,n}\,u_{m,n}\,,
	\qquad
	f = \sum_{m,n}f_{m,n}\,u_{m,n}\,,
	\label{eq:expansion}
\end{equation}
the gap conditions are enforced by construction: $\chi|_\mathrm{gap}=0$
implies $J_\phi|_\mathrm{gap}=0$ at the toroidal gaps, and
$\partial_{\phi'}\chi|_{\theta=\pi}=0$ implies $J_\theta=0$ at the
poloidal gap. Substituting \eqref{eq:expansion} into
\eqref{eq:poisson} and using \eqref{eq:eigenfunctions}, the Poisson
equation reduces mode by mode to an algebraic relation:
\begin{equation}
	\chi_{m,n} = -\frac{f_{m,n}}{k_m^2+k_n^2}\,.
	\label{eq:chi_modal_solution}
\end{equation}
The dependence on $N_g$ enters only through the denominator
$k_m^2+k_n^2$: as $N_g$ increases, $k_n$ grows and the modal
amplitudes are suppressed.

For a purely toroidal applied field ($E_\theta=0$), assumed
toroidally uniform, $E_\phi=V_\phi^\mathrm{tot}/(2\pi R_0)$, the
source reduces to
\begin{equation}
	f=-\frac{\sigma t_w E_\phi\sin\theta}{h_\phi}\,,
	\label{eq:source}
\end{equation}
which is independent of $N_g$. The numerical solver computes the
coefficients $f_{m,n}$ of the exact source \eqref{eq:source} via an
FFT in $\theta$ and a DST-I in $\phi'$, applies
\eqref{eq:chi_modal_solution} to all modes, and obtains the current
densities from \eqref{eq:J_from_chi_def} by inverse transformation.

Insight into the structure of the solution is gained at leading
order in $r_w/R_0$. In the cylindrical approximation $h_\phi\approx R_0$,
\eqref{eq:source} becomes $f\approx A\sin\theta$, with
$A=-\sigma t_w E_\phi/R_0$: the source contains only the poloidal mode
$m=1$, while the exact metric adds harmonics $m\geq2$ suppressed by
successive factors of $r_w/R_0\approx0.26$, which are responsible for
the outboard--inboard asymmetry discussed in
Sec.~\ref{sec:res_currents}. Since the source is constant in $\phi'$,
its DST-I coefficients are
\begin{equation}
	f_{1,n} = \frac{2}{L_s}\int_0^{L_s} A\,
	\sin\!\left(\frac{n\pi\phi'}{L_s}\right)\mathrm{d}\phi'
	= \frac{4A}{n\pi}\quad(n\ \mathrm{odd})\,,
	\qquad f_{1,n}=0\quad(n\ \mathrm{even})\,,
	\label{eq:f_dst}
\end{equation}
and \eqref{eq:chi_modal_solution} gives the separable stream function
$\chi(\theta,\phi')=\sin\theta\cdot g(\phi')$, with
\begin{equation}
	g(\phi') = -\sum_{n\ \mathrm{odd}}^{N_\mathrm{int}}
	\frac{4A}{n\pi\,(k_1^2+k_n^2)}\,
	\sin\!\left(\frac{n\pi\phi'}{L_s}\right),
	\label{eq:g_def}
\end{equation}
and current densities
\begin{equation}
	J_\theta = \frac{\sin\theta}{h_\phi}\,g'(\phi')\,,\qquad
	J_\phi = -\frac{\cos\theta}{r_w}\,g(\phi')\,.
	\label{eq:J_from_chi}
\end{equation}
This leading-order
form captures the dominant structure of the numerical solution and,
since $m=1$ also dominates the MHD spectrum of an RFP, it is the
basis for the interpretation of the results and of the scaling laws
of Sec.~\ref{sec:scaling}.

Once the eddy current distribution is determined, the magnetic field
at the plasma surface ($r_\mathrm{pla}=0.490\,\mathrm{m}$) is
computed by summing the Biot-Savart contributions from all sectors:
\begin{equation}
	\mathbf{B}(\mathbf{r}) = \frac{\mu_0}{4\pi}
	\sum_{k=0}^{N_g-1}
	\int \frac{\mathbf{J}_k(\mathbf{r}') \times (\mathbf{r}-\mathbf{r}')}
	{|\mathbf{r}-\mathbf{r}'|^3}\,\mathrm{d}S'\,,
	\label{eq:savart}
\end{equation}
where $\mathbf{J}_k$ is the surface current density of the $k$-th
sector, obtained by rotating the reference solution by $2\pi k/N_g$
about the vertical axis.

\section{Results}
\label{sec:res}

The results are presented for the ramp-up phase of shot \#29262 at
$t=20\,\mathrm{ms}$, with $V_\phi^\mathrm{tot}=-99.0\,\mathrm{V}$.
The 2-gap configuration is discussed first in detail; the scaling with
$N_g$ is analyzed for $N_g = 1, 2, 4, 8, 10, 20, 50, 100, 200$.

\subsection{Eddy current distribution in the 2-gap configuration}
\label{sec:res_currents}

\begin{figure}[tb!]
	\centering
	\includegraphics[width=.495\textwidth]{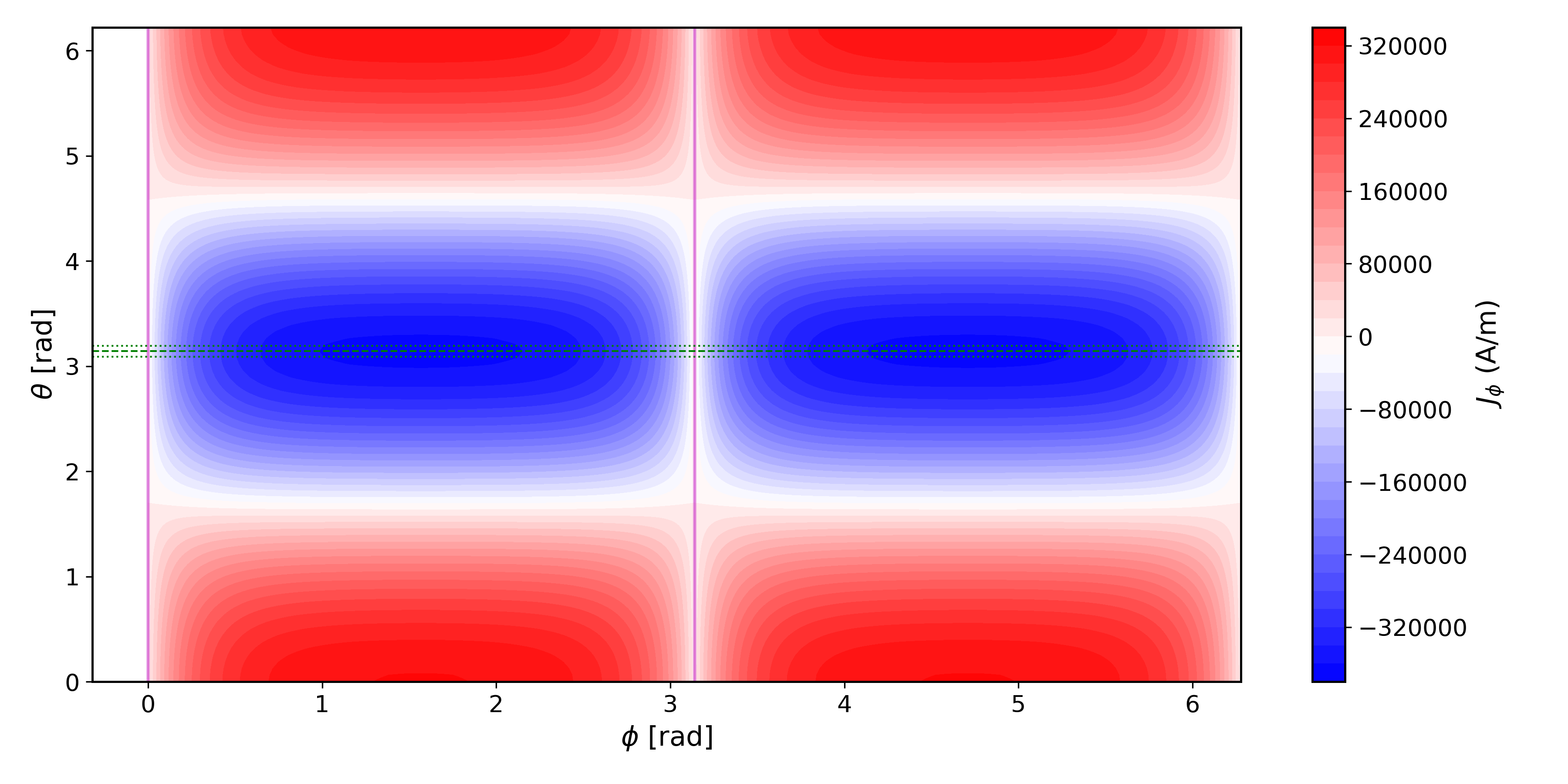}
	\includegraphics[width=.495\textwidth]{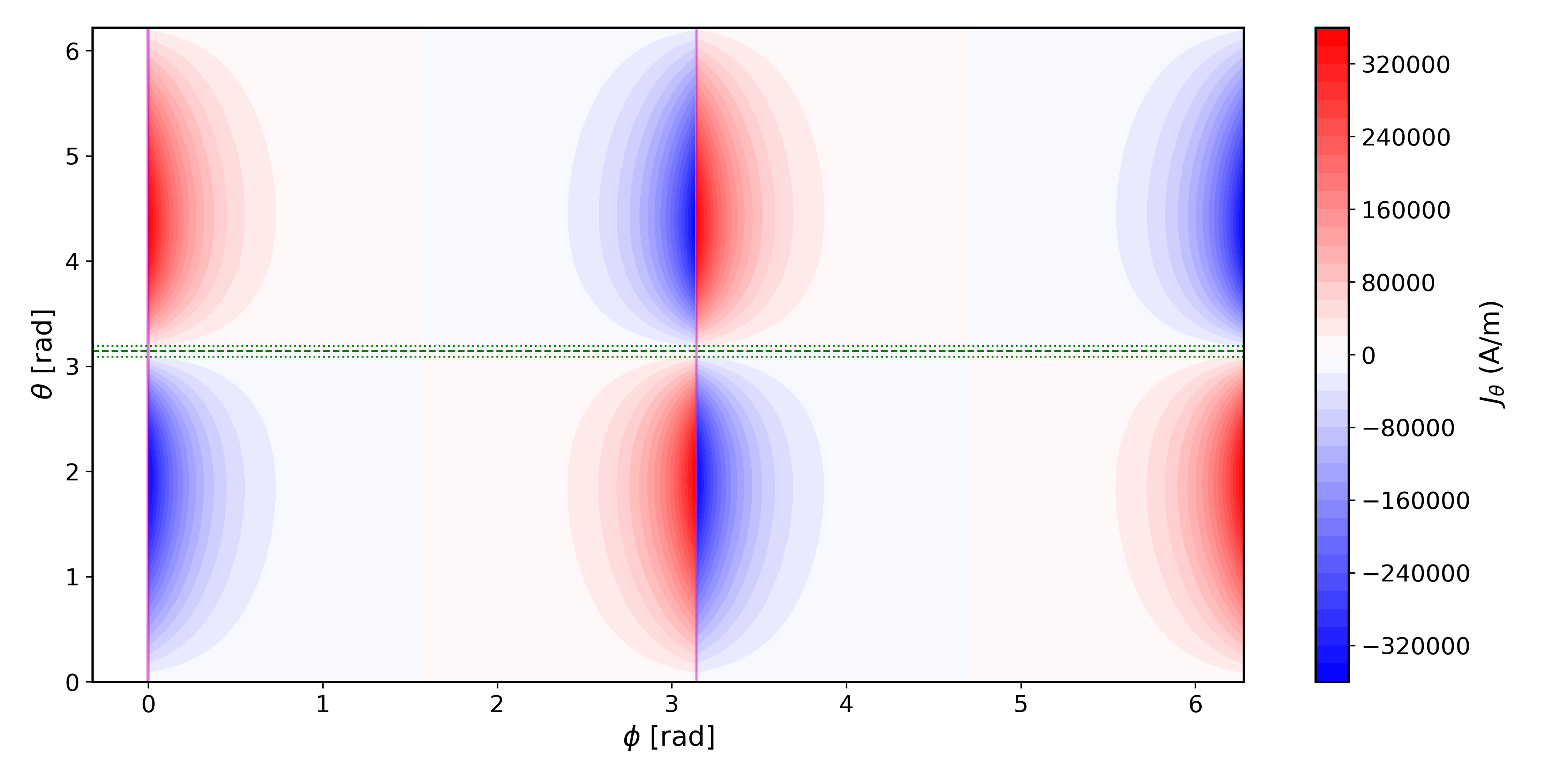}
	\caption{2-gap configuration: 2D maps of $J_\phi$ (left) and
		$J_\theta$ (right).}
	\label{fig:J_behaviour_2d}
\end{figure}

\begin{figure}[tb!]
	\centering
	\includegraphics[width=.7\textwidth]{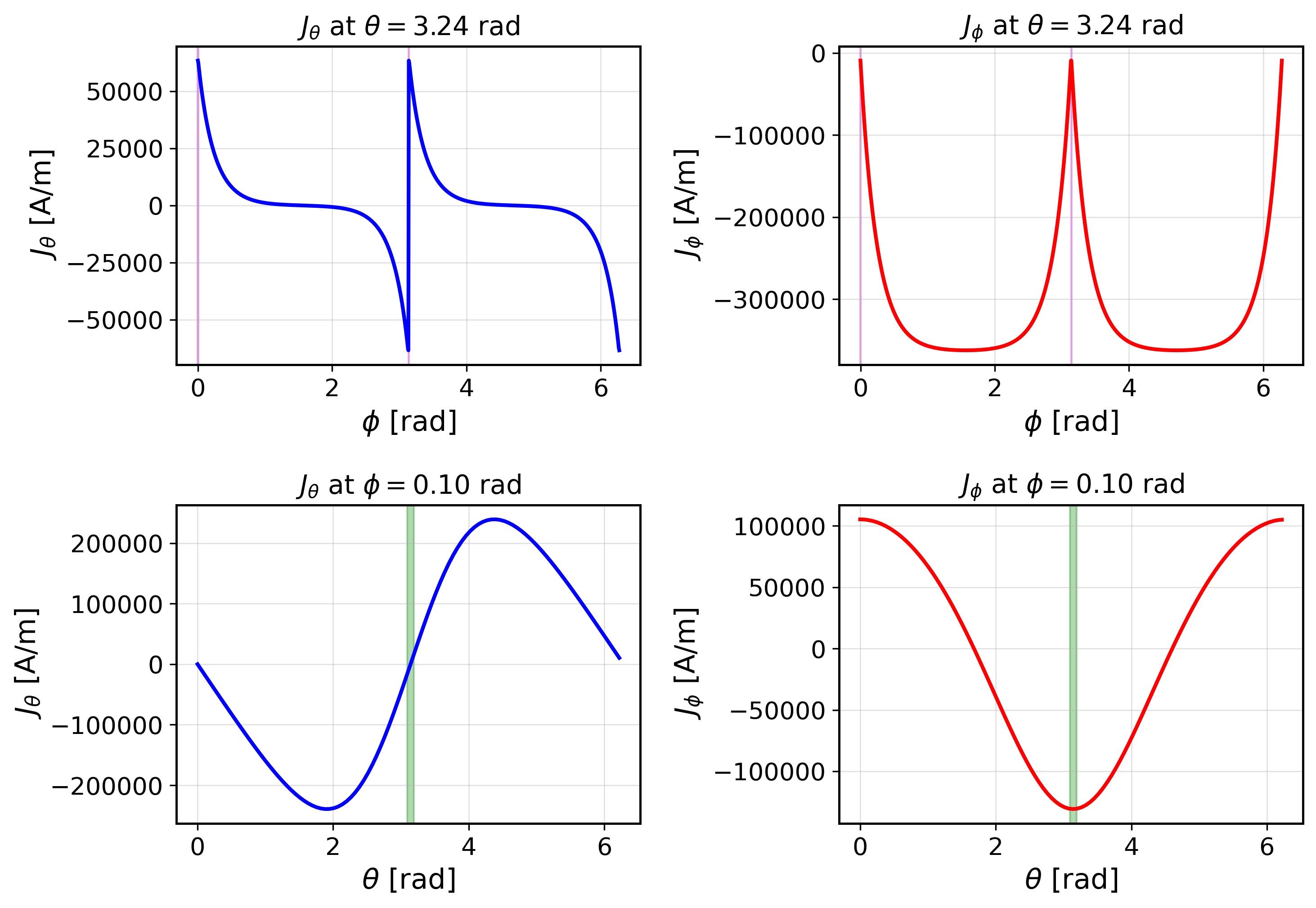}
	\caption{2-gap configuration: poloidal profiles of $J_\phi$ and
		$J_\theta$ close to the gaps.}
	\label{fig:J_behaviour_2gap}
\end{figure}

\begin{figure}[tb!]
	\centering
	\includegraphics[width=.8\textwidth]{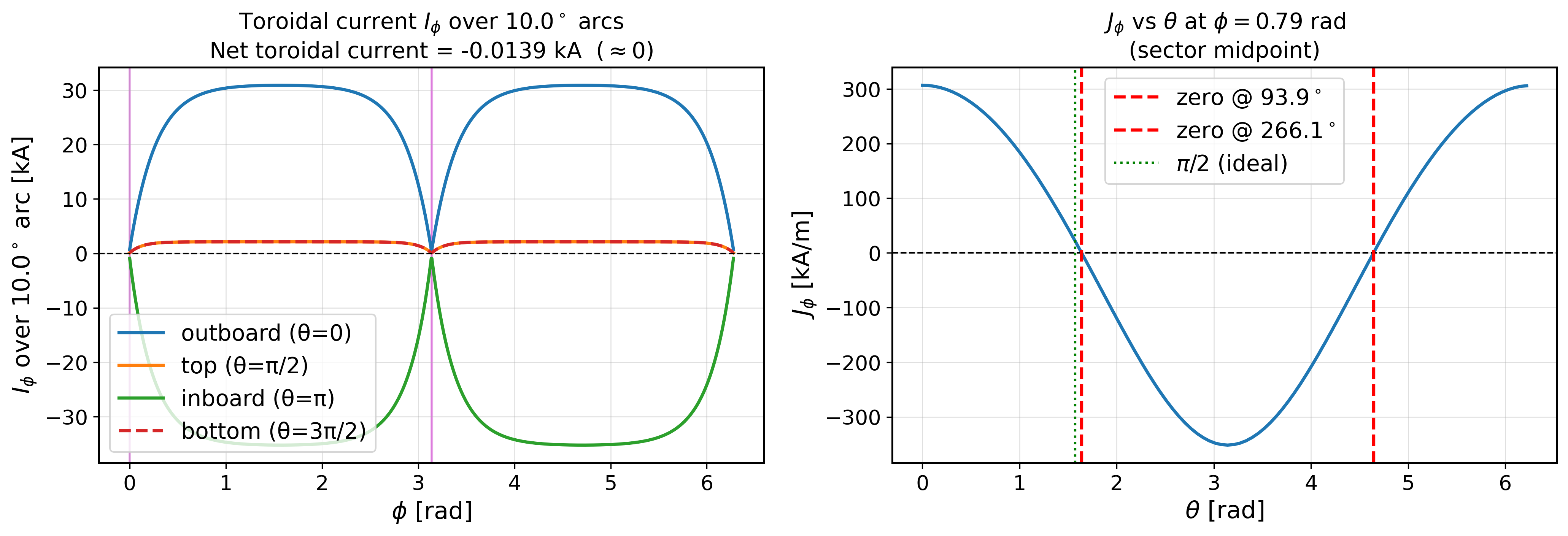}
	\includegraphics[width=.8\textwidth]{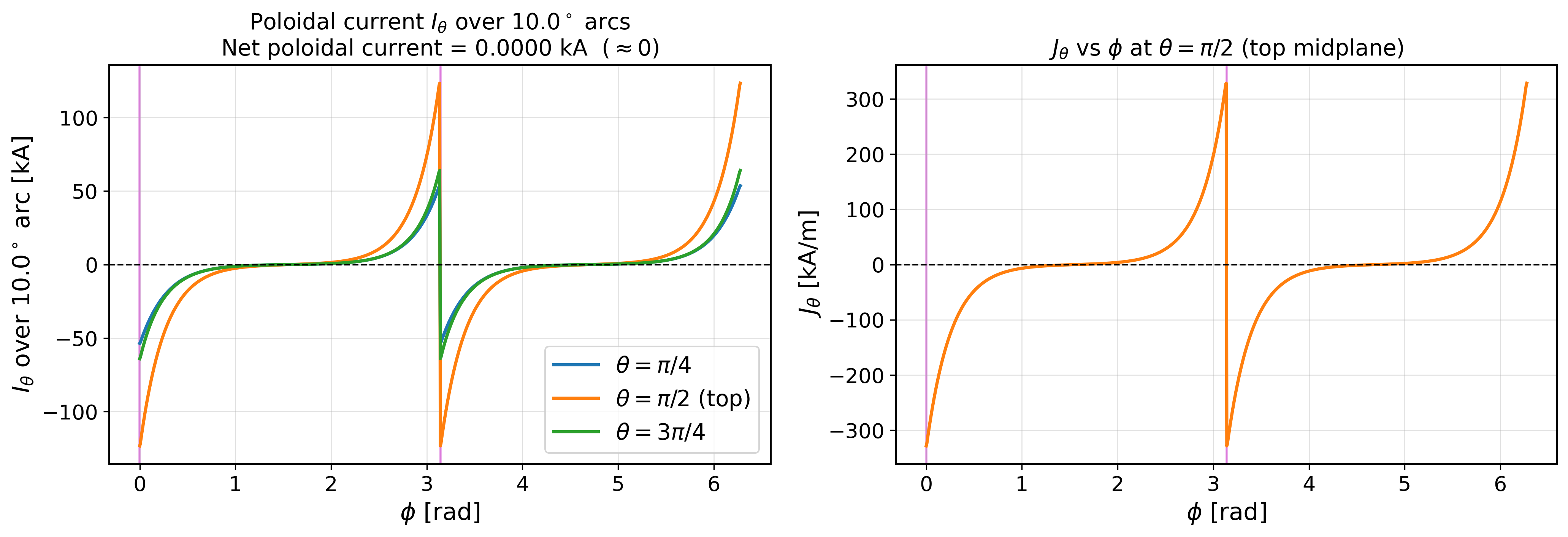}
	\caption{2-gap configuration. Top: integrated toroidal current
		$I_\phi$ at four poloidal positions (left) and $J_\phi(\theta)$
		at the sector midpoint, showing the zero-crossing shift to
		$93.9^\circ$ (right). Bottom: poloidal current $I_\theta$ crossing a $10^\circ$ toroidal arc
		at three poloidal positions (left) and $J_\theta(\phi)$ at
		$\theta=\pi/2$ (right).}
	\label{fig:I_total}
\end{figure}

The 2D distributions of $J_\phi$ and $J_\theta$ are shown in
Fig.~\ref{fig:J_behaviour_2d} and Fig.~\ref{fig:J_behaviour_2gap}. The toroidal current density $J_\phi$
has a dominant $\cos\theta$ poloidal distribution, consistent with the leading-order solution \eqref{eq:J_from_chi}. Due to
the toroidal geometry ($h_\phi = R_0 + r_w\cos\theta$), the
zero-crossing of $J_\phi$ is shifted from the ideal $\theta=\pi/2$
to $93.9^\circ$ — a shift of $\approx4^\circ$ (Fig. \ref{fig:I_total}) due to the
outboard–inboard asymmetry of $h_\phi$ — and the peaks at $\theta=0$
and $\theta=\pi$ differ in amplitude for the same reason. The
dependence on $\phi$ is governed by the gap boundary condition:
$J_\phi$ vanishes exactly at the gap locations. The poloidal current density $J_\theta$ follows a $\sin\theta$
distribution and vanishes at both midplanes ($\theta=0$ and
$\theta=\pi$), as predicted by \eqref{eq:J_from_chi} and expected from the $m=1$ mode structure. It is
concentrated near the gaps: starting from its maximum at the gap
edge, it decreases by a factor $e$ over a toroidal angle
$\Delta\phi\approx0.24\,\mathrm{rad}$
(Fig.~\ref{fig:J_behaviour_2gap}, top-left), corresponding to an arc
$R_0\Delta\phi\approx0.48\,\mathrm{m}$, close to the shell radius
$r_w=0.51\,\mathrm{m}$, and it vanishes at the sector midpoint by
symmetry. The return current thus flows in a layer of width
$\sim r_w$ around each gap.

The toroidal current crossing a poloidal arc of width
$\Delta\theta=10^\circ$ centred at $\theta$, at fixed toroidal angle $\phi$, is
\begin{equation}
	I_\phi(\theta,\phi) =
	\int_{\theta-\Delta\theta/2}^{\theta+\Delta\theta/2}
	J_\phi(\theta',\phi)\,r_w\,\mathrm{d}\theta'
	\quad [\mathrm{A}]
\end{equation}
and reaches $\pm30\,\mathrm{kA}$ at the outboard and inboard
midplanes, while it is negligible at the top and bottom, consistent
with the $\cos\theta$ distribution of $J_\phi$
(Fig.~\ref{fig:I_total}, top). The net toroidal current through any
poloidal cross-section vanishes identically,
$\int_0^{2\pi}J_\phi\,r_w\,\mathrm{d}\theta =
-[\chi(2\pi,\phi')-\chi(0,\phi')]=0$, by periodicity of $\chi$ in
$\theta$; numerically, the residual is $\sim14\,\mathrm{A}$, negligible
compared to the arc current of $\sim30\,\mathrm{kA}$. The pointwise
divergence is $|\nabla\cdot\mathbf{J}|_\mathrm{max} =
7.4\times10^{-9}\,\mathrm{A/m^2}$ (mean
$8.0\times10^{-10}\,\mathrm{A/m^2}$), confirming that the discrete
operators preserve the solenoidal condition. The poloidal current crossing a toroidal arc of width
$\Delta\phi=10^\circ$ centred at $\phi$, at fixed poloidal angle
$\theta$, is
\begin{equation}
	I_\theta(\theta,\phi) =
	\int_{\phi-\Delta\phi/2}^{\phi+\Delta\phi/2}
	J_\theta(\theta,\phi')\,h_\phi(\theta)\,\mathrm{d}\phi'
	\quad [\mathrm{A}]
\end{equation}
and peaks near the gap locations (Fig.~\ref{fig:I_total},
bottom-left), where the toroidal current must redirect poloidally to
satisfy $J_\phi|_\mathrm{gap}=0$. The net poloidal current through
each sector vanishes identically,
$\int_0^{L_s} J_\theta h_\phi\,\mathrm{d}\phi' =
\chi(\theta,L_s)-\chi(\theta,0)=0$, so that the return current forms
closed loops within each sector, consistent with
$\nabla\cdot\mathbf{J}=0$.

\subsection{Magnetic fields in the 2-gap configuration}
\label{sec:res_Br}

\begin{figure}[tb!]
	\centering
	\includegraphics[width=.9\textwidth]{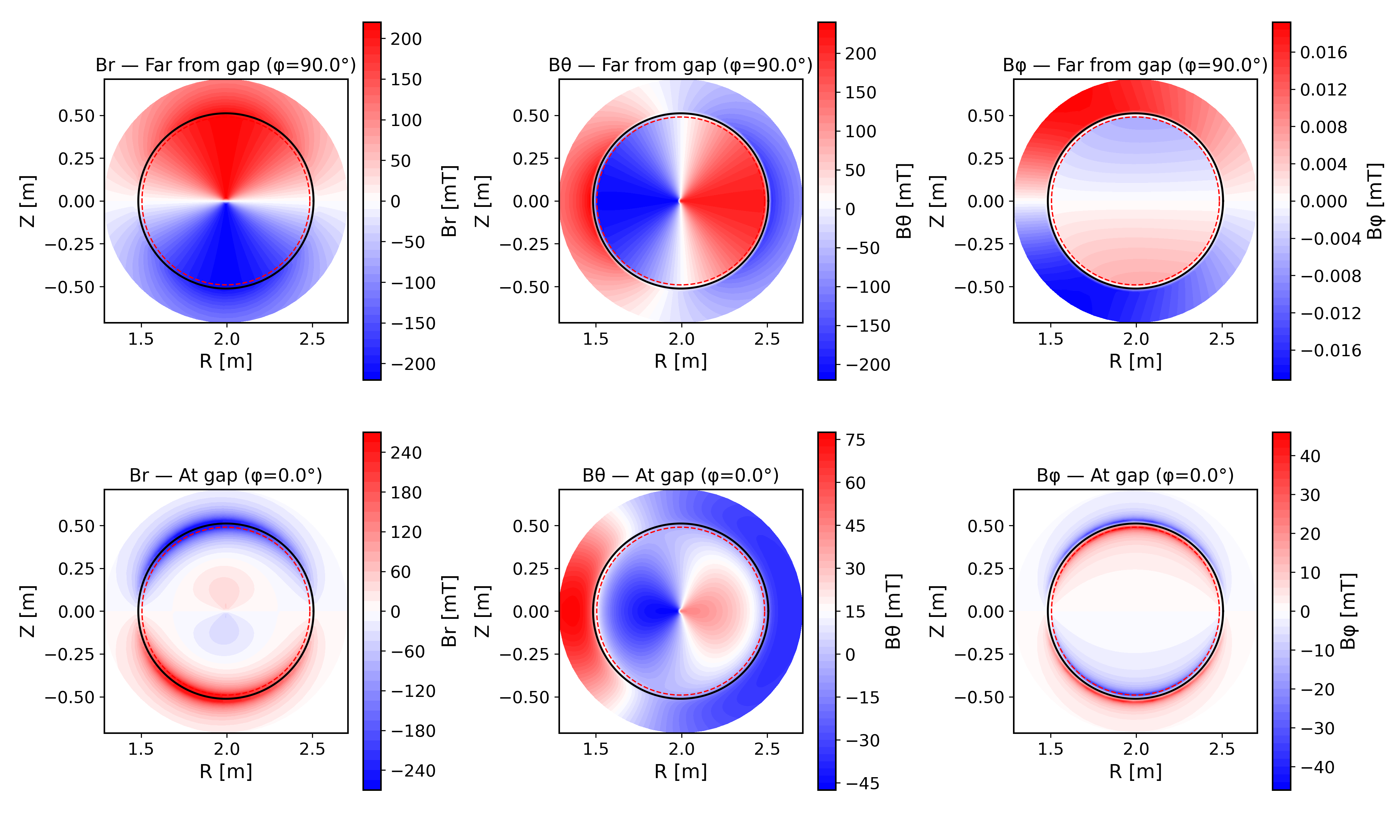}
	\caption{2-gap configuration: radial, poloidal and toroidal
		magnetic field components on a poloidal cross section grid, far
		from the gap (top row) and at the gap (bottom row).}
	\label{fig:Bgrid_2gaps}
\end{figure}

\begin{figure}[h]
	\centering
	\includegraphics[width=1\textwidth]{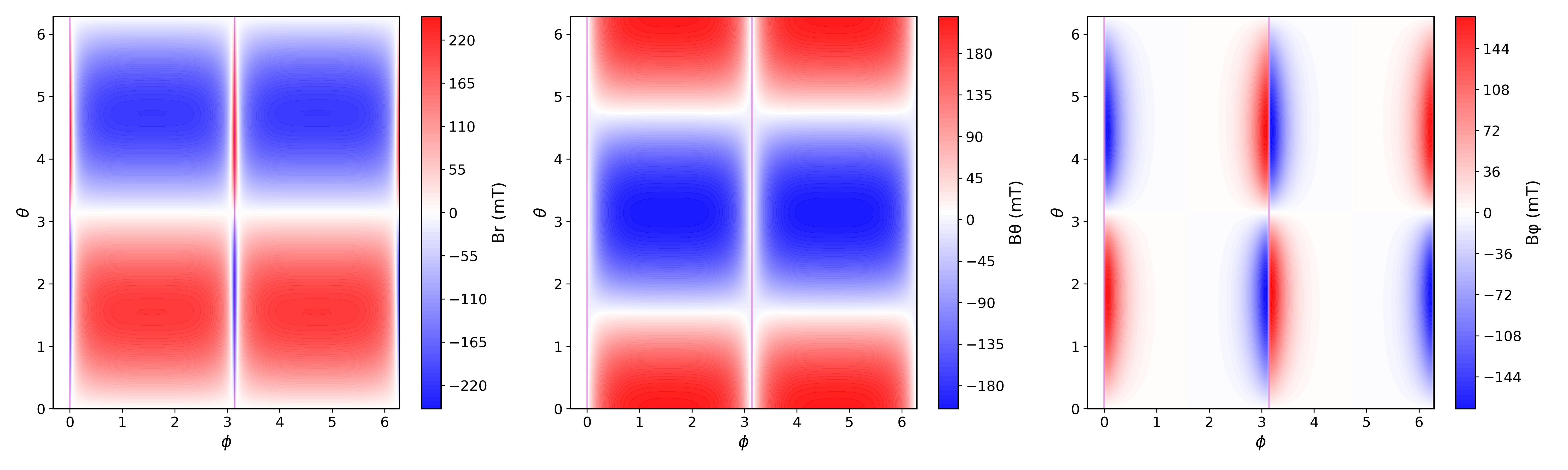}
	\includegraphics[width=.6\textwidth]{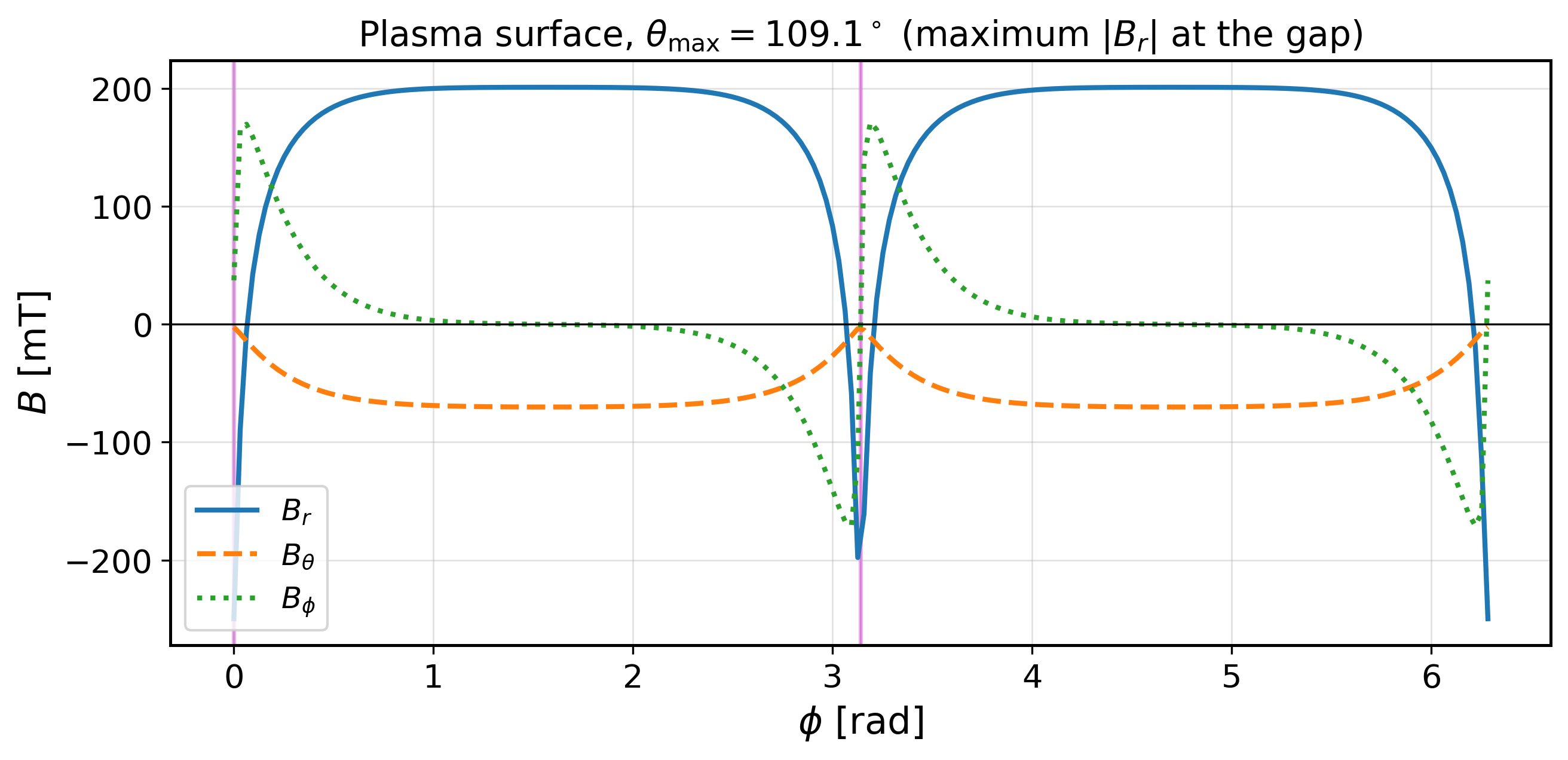}
\caption{2-gap configuration: magnetic field components at the plasma
	surface (top) and their toroidal profiles at
	$\theta_\mathrm{max}=109^\circ$, where $|B_r|$ at the gap is
	maximum (bottom).}
	\label{fig:Bpla_2gaps}
\end{figure}

The eddy current magnetic field components computed on a poloidal cross section and at the
plasma surface ($r=0.490\,\mathrm{m}$) are shown in
Figs.~\ref{fig:Bgrid_2gaps} and~\ref{fig:Bpla_2gaps}, respectively. Far from the
gap ($\phi=90°$, sector midpoint), $B_r$ reaches
$\approx215\,\mathrm{mT}$ and exhibits a clean dipolar pattern in the
poloidal cross-section, consistent with the $\cos\theta$ distribution
of $J_\phi$. The value at the plasma surface ($\theta=109°$, Fig. \ref{fig:Bpla_2gaps}) is
$B_r^\mathrm{far}\sim201\,\mathrm{mT}$, in agreement with
the CARIDDI results \cite{eps2026}

At the gap ($\phi=0°$), the radial field reverses sign and increases
in magnitude: $B_r^\mathrm{gap}\sim-250\,\mathrm{mT}$ at the plasma
surface, giving $|B_r^\mathrm{gap}/B_r^\mathrm{far}|=1.24$ and a
total $B_r$ excursion across the gap of about twice the far-field
value. This sign reversal is the direct consequence of the poloidal
return current $J_\theta$ concentrating at the gap to satisfy
$J_\phi|_\mathrm{gap}=0$ and $\nabla\cdot\mathbf{J}=0$: the current
loop formed by $J_\theta$ produces a radial field of opposite sign to
the bulk toroidal current contribution. The effect is visible as two
opposite values in the poloidal cross-section
(Fig.~\ref{fig:Bgrid_2gaps}, bottom-left) and as a sharp drop in the
toroidal profile (Fig.~\ref{fig:Bpla_2gaps}, bottom). 
The maximum of $|B_r^\mathrm{gap}|$ occurs at
$\theta_\mathrm{max}=109^\circ$ rather than $90^\circ$, consistent
with the asymmetric shift of $J_\phi$. The sign reversal and its relative magnitude are in agreement with previous results \cite{masiello_analytical_1998,marchioritn,eps2026}, confirming that the DST solver correctly captures the gap physics. The toroidal component $B_\phi$ is negligible far from the gap ($\lesssim0.016\,\mathrm{mT}$) but
reaches $\approx40\,\mathrm{mT}$ at the gap, produced by $J_\theta$.
The poloidal component $B_\theta$ is
approximately toroidally uniform and weakly perturbed at the gap (Fig. \ref{fig:Bpla_2gaps}).

\subsection{Scaling with number of gaps}
\label{sec:scaling}

\begin{figure}[h!]
	\centering
	\includegraphics[width=.9\textwidth]{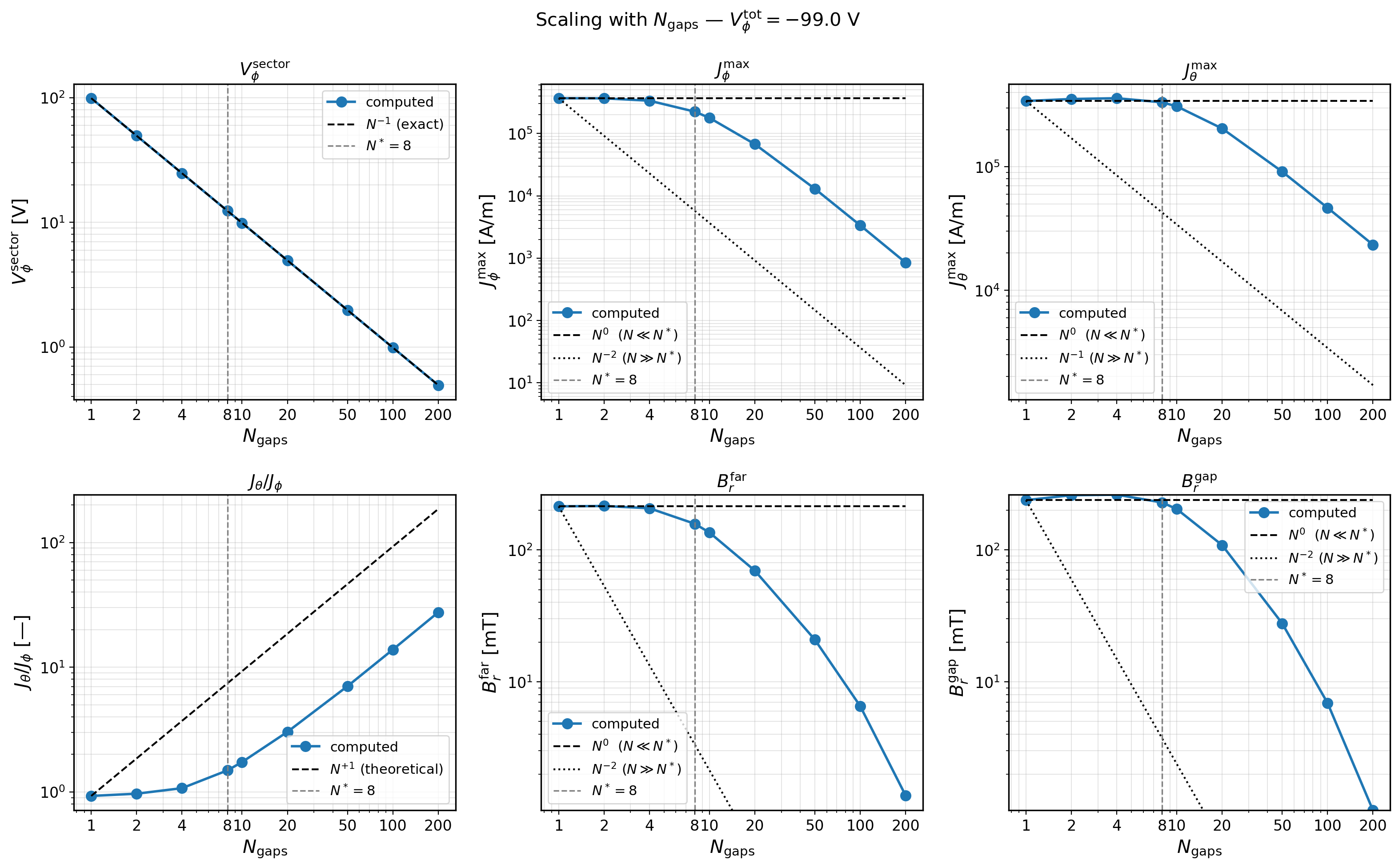}
	\caption{Scaling of $V_\phi^\mathrm{sector}$, $J_\phi^\mathrm{max}$,
		$J_\theta^\mathrm{max}$, $J_\theta/J_\phi$,
		$B_r^\mathrm{far}$ and $B_r^\mathrm{gap}$ with $N_g$
		($N_g=1$--$200$). Black lines: asymptotic dependencies
		\eqref{eq:Jphi_scaling}--\eqref{eq:Brgap_scaling}. Vertical
		grey line: $N^*\approx8$.}
	\label{fig:scaling_Ngaps}
\end{figure}

The radial field has two physically distinct contributions. Far from
the gap, $B_r^\mathrm{far}$ is produced by the toroidally uniform
component of $J_\phi$, i.e.\ the axisymmetric ($n=0$) part of the
eddy current. At the gap, $J_\theta$ produces an additional field of
opposite sign, breaking toroidal symmetry (Fig.~\ref{fig:Bpla_2gaps}).
The error field is quantified as the excess of the radial field
magnitude at the gap over its far-from-the-gap value:
\begin{equation}
	B_{ef} = |B_r^\mathrm{gap}| - |B_r^\mathrm{far}|\,,
	\label{eq:Bef}
\end{equation}
which vanishes for a continuous shell. Results for all $N_g$ values
are summarised in Tab.~\ref{tab:scaling_Ngaps} and
Fig.~\ref{fig:scaling_Ngaps}.

\begin{table}[tb]
	\centering
	\caption{Scaling of induced currents and radial field with number
		of gaps (shot \#29262, $t=20\,\mathrm{ms}$).
		$B_r^\mathrm{far}$ and $B_r^\mathrm{gap}$ are the maxima of
		$|B_r|$ over $\theta$ at the plasma surface, at the sector
		midpoint and at the gap centre, respectively.}
	\label{tab:scaling_Ngaps}
	\begin{tabular}{c|c|c|c|c|c|c}
		$N_g$ & $V_\phi^\mathrm{sector}$ (V) & $J_\phi^\mathrm{max}$
		(kA/m) & $J_\theta^\mathrm{max}$ (kA/m) &
		$B_r^\mathrm{far}$ (mT) & $B_r^\mathrm{gap}$ (mT) &
		$B_{ef}$ (mT) \\
		1   & 99.0 & 367 & 341 & 214 & 238 & 24 \\
		2   & 49.5 & 366 & 354 & 215 & 260 & 45 \\
		4   & 24.8 & 335 & 359 & 207 & 262 & 55 \\
		8   & 12.4 & 224 & 332 & 158 & 229 & 71 \\
		10  &  9.9 & 178 & 309 & 136 & 204 & 68 \\
		20  &  5.0 &  67 & 204 &  70 & 109 & 39 \\
		50  &  2.0 &  13 &  91 &  21 &  28 &  7 \\
		100 &  1.0 &   3 &  47 &   7 &   7 &  0 \\
		200 &  0.5 &   1 &  23 &   1 &   1 &  0 \\
	\end{tabular}
\end{table}

The scaling is governed by the competition between the poloidal
wavenumber $k_1=1/r_w$ and the toroidal segmentation wavenumber
$k_s=N_g/(2R_0)$, which enter the leading-order solution
\eqref{eq:g_def} through the denominator $k_1^2+n^2k_s^2$ in \eqref{eq:chi_modal_solution}. The
same competition arises in the effective resistivity of the
Armadillo wall-time model~\cite{walltimearmadillo2026}. Written in
terms of the poloidal, helical and segmentation wavenumbers $m/r_w$,
$n/R_0$ and $k_s$, it reads
\begin{equation}
	\eta_\mathrm{eff}(m,n;N_g)
	= \frac{(m/r_w)^2\,\eta_\phi + (n/R_0)^2\,\eta_\theta
		+ k_s^2\,\eta_\theta}{(m/r_w)^2+(n/R_0)^2}\,,
	\label{eq:eta_eff_general}
\end{equation}
which for $(m=1,\,n=0)$ and isotropic resistivity reduces to
\begin{equation}
	\frac{\eta_\mathrm{eff}(N_g)}{\eta}
	= \frac{k_1^2+k_s^2}{k_1^2}
	= 1+\left(\frac{r_wN_g}{2R_0}\right)^2,
	\label{eq:eta_eff_simple}
\end{equation}
i.e.\ the Poisson denominator of the fundamental DST mode ($n=1$)
normalised to its continuous-shell value. The poloidal and
segmentation contributions are equal when
\begin{equation}
	k_s = k_1
	\quad\Longleftrightarrow\quad
	\frac{N_g}{2R_0} = \frac{1}{r_w}
	\quad\Longrightarrow\quad
	N^* = \frac{2R_0}{r_w}\approx 8
	\label{eq:Nstar}
\end{equation}
for RFX-mod2, marked by the vertical grey line in
Fig.~\ref{fig:scaling_Ngaps}. The asymptotic scaling laws are:
\begin{align}
	J_\phi &\sim
	\begin{cases} N_g^{0}  & N_g \ll N^* \\
		N_g^{-2} & N_g \gg N^* \end{cases}
	\label{eq:Jphi_scaling}\\[4pt]
	J_\theta &\sim
	\begin{cases} N_g^{0}  & N_g \ll N^* \\
		N_g^{-1} & N_g \gg N^* \end{cases}
	\label{eq:Jtheta_scaling}\\[4pt]
	J_\theta/J_\phi &\sim
	\begin{cases} 1   & N_g \ll N^* \\
		N_g & N_g \gg N^* \end{cases}
	\label{eq:ratio_scaling}
\end{align}
For $N_g\ll N^*$ the gaps affect the current only within a toroidal
distance of order $r_w$ from each gap (Sec.~\ref{sec:res_currents}),
where $J_\phi$ drops to zero and the current turns poloidally: the return current $J_\theta$ is
confined in a layer of width $\sim r_w$, independent of the sector
length, and both current components are independent of $N_g$, as in
a continuous shell. For $N_g\gg N^*$ the half-length of each sector, $\pi R_0/N_g$, becomes
smaller than $r_w$: the layers adjacent to the two gaps of a sector
overlap, and $J_\phi$, which must vanish at both ends of a short
sector, can no longer reach its continuous-shell value. Its maximum
then decreases with the square of the sector length
($\propto N_g^{-2}$), while $J_\theta$, set by the variation of the
stream function along the sector, decreases with the sector length
itself ($\propto N_g^{-1}$).

These dependencies are shown as black lines in
Fig.~\ref{fig:scaling_Ngaps}. The voltage per sector (top-left panel)
follows the exact $N_g^{-1}$ law, while the driving field $E_\phi$
remains constant. The toroidal current (top-centre panel) shows a
plateau of $\approx350\,\mathrm{kA/m}$ for $N_g\leq4$ and the
$N_g^{-2}$ decay for $N_g\geq20$. The poloidal current (top-right
panel) remains between $341$ and $359\,\mathrm{kA/m}$ for $N_g\leq8$
and decays as $N_g^{-1}$ for $N_g\geq20$; the slight maximum at
$N_g\simeq4$ is a toroidal correction. The ratio $J_\theta/J_\phi$
(bottom-left panel) is close to unity for $N_g\leq4$ and grows
linearly with $N_g$ beyond $N^*$: in the segmented regime the shell
carries predominantly poloidal current, flowing in closed loops
around each gap.

The radial field scaling follows from the Biot-Savart law
\eqref{eq:savart}. Far from the gap, $B_r^\mathrm{far}$ is produced
by $J_\phi$ distributed over the sector. At the gap,
$B_r^\mathrm{gap}$ is produced by the return-current layer, whose
toroidal width is $\sim r_w$ for $N_g\ll N^*$ and $\sim\pi R_0/N_g$
for $N_g\gg N^*$. Hence:
\begin{align}
	B_r^\mathrm{far} &\sim
	\begin{cases} N_g^{0}  & N_g \ll N^* \\
		N_g^{-2} & N_g \gg N^* \end{cases}
	\label{eq:Brfar_scaling}\\[4pt]
	B_r^\mathrm{gap} &\sim
	\begin{cases} N_g^{0}  & N_g \ll N^* \\
		N_g^{-2} & N_g \gg N^* \end{cases}
	\label{eq:Brgap_scaling}
\end{align}
Both behaviours are confirmed in the bottom-centre and bottom-right
panels of Fig.~\ref{fig:scaling_Ngaps}: $B_r^\mathrm{far}$ shows a
plateau at $\approx215\,\mathrm{mT}$ followed by a decay down to
$\approx1\,\mathrm{mT}$ at $N_g=200$, while $|B_r^\mathrm{gap}|$
remains above $|B_r^\mathrm{far}|$ for $N_g\lesssim N^*$ and
converges to it for $N_g\gtrsim100$, where the toroidal modulation
of $B_r$ at the plasma surface vanishes.

\begin{figure}[tb]
	\centering
	\includegraphics[width=0.4\textwidth]{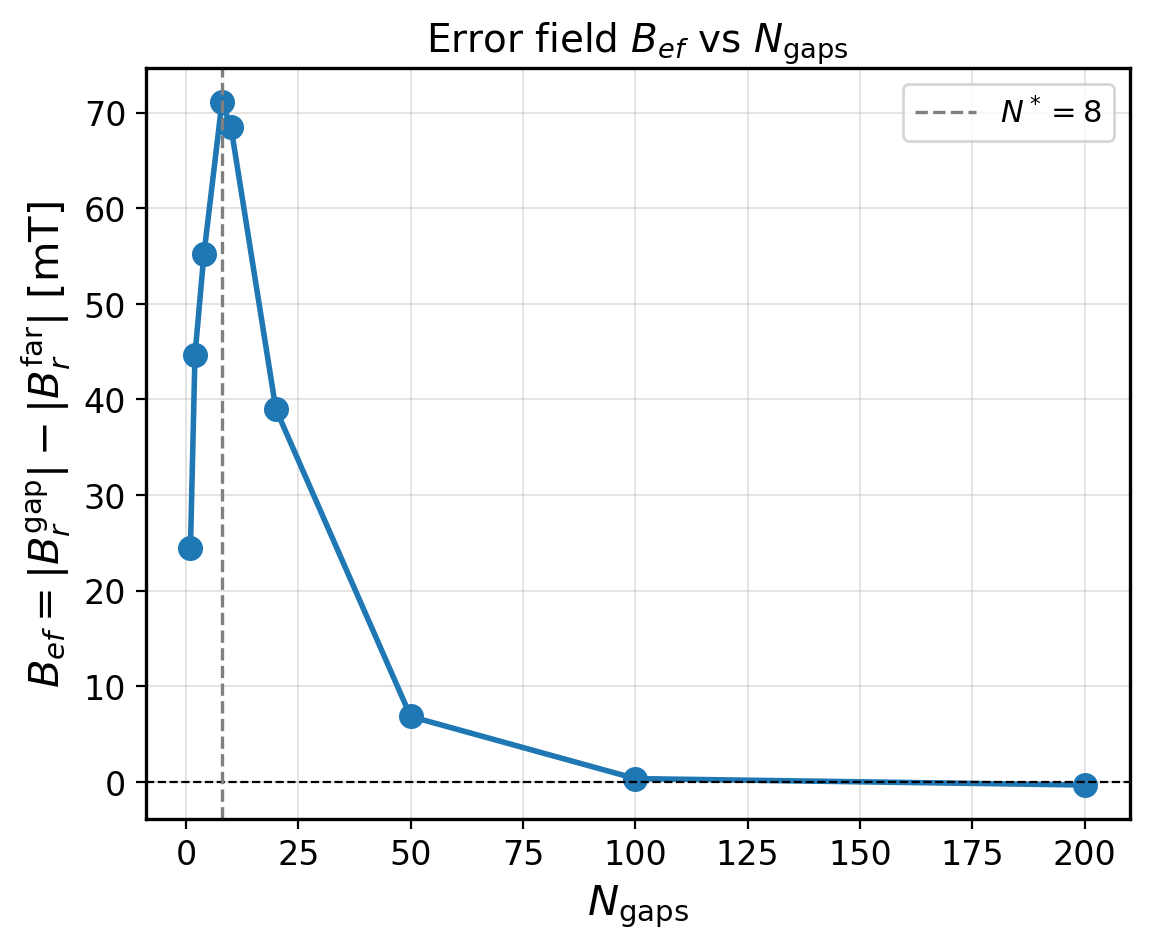}
	\caption{Error field $B_{ef}=|B_r^\mathrm{gap}|-|B_r^\mathrm{far}|$
		as a function of $N_g$. The maximum occurs near $N^*\approx8$;
		for $N_g\gg N^*$ the error field decays as $N_g^{-2}$.}
	\label{fig:Bef_scaling}
\end{figure}

The error field $B_{ef}$ (Fig.~\ref{fig:Bef_scaling}) grows from
$24\,\mathrm{mT}$ at $N_g=1$ to a maximum of $\approx71\,\mathrm{mT}$
at $N_g\simeq N^*\approx8$, and then decays as $N_g^{-2}$, falling
below $1\,\mathrm{mT}$ for $N_g\gtrsim100$. The maximum occurs near
$N^*$ because, approaching the transition, $J_\phi$ starts to
decrease before $J_\theta$: the far field drops while the gap field
is still nearly unchanged, and the difference of their magnitudes
grows. Beyond $N^*$ both fields are suppressed together. Since
$N^*=2R_0/r_w$ depends only on the aspect ratio, the transition scale
is a purely geometric quantity: lower aspect ratio devices are more
sensitive to segmentation, while higher aspect ratio devices are more
resilient. The scaling of $B_{ef}$ with $N_g$ has direct implications
for MHD stability, affecting both resistive wall mode stabilisation
and tearing mode activity.

\section{Conclusions}
\label{sec:conclusions}

A stream function formulation for the eddy-current
distribution in a toroidally segmented conducting shell has been
developed and verified. The
Poisson-like equation for the stream function is solved spectrally via a
Discrete Sine Transform, which enforces the zero-current boundary
condition at each gap exactly by construction, and the
magnetic field at the plasma surface is obtained by Biot-Savart
integration of the full-torus current distribution. Applied to the
ramp-up phase of RFX-mod2 with two gaps, the method reproduces the
sign reversal of the radial field at the gap and its magnitude found
in previous analytical and 3D electromagnetic calculations.

The toroidal current $J_\phi$ flows in the same direction on every
sector with a dominant $\cos\theta$ poloidal profile, while the
poloidal return current $J_\theta$ is concentrated in a layer of toroidal
width $\sim r_w$ around each gap and reverses sign between adjacent
sectors.  Their scaling with $N_g$ is governed
by the competition between the poloidal geometric scale $k_1=1/r_w$
and the segmentation scale $k_s=N_g/(2R_0)$, with a transition at
$N^*=2R_0/r_w\approx8$ for RFX-mod2, in agreement with
the effective-resistivity model of~\cite{walltimearmadillo2026}. At
the gap, $J_\theta$ produces a radial field of opposite sign to the value far from the gap; the resulting error field $B_{ef}$ reaches its maximum near
$N_g=N^*$ and decays as $N_g^{-2}$ for
$N_g\gg N^*$, becoming negligible for $N_g\gtrsim100$. These results provide the
essential input for assessing the impact of Armadillo segmentation
on MHD stability, in particular on resistive wall mode stabilisation
and tearing mode activity.

\bibliographystyle{unsrt}
\bibliography{refs}

\end{document}